\documentclass[letterpaper, onecolumn, 11pt]{article}
\usepackage[left=2cm,right=2cm,top=2cm,bottom=2cm]{geometry}

\usepackage[utf8x]{inputenc}
\usepackage{url}
\usepackage{microtype}
\usepackage{graphicx}
\usepackage{booktabs}
\usepackage{tabulary}
\usepackage{subfigure}

\usepackage{setspace}

\usepackage[unicode=true]{hyperref}

\hypersetup{breaklinks=true,
            bookmarks=true,
            pdfauthor={},
            pdftitle={},
            colorlinks=true,
            urlcolor=blue,
            linkcolor=magenta,
            pdfborder={0 0 0}}

\title{Zahir: a Object-Oriented Framework for Computer Graphics}

\author{Eduardo Graells-Garrido
        and Mar\'ia Cecilia Rivara
        \\
         Department of Computer Science  \\ University of Chile  \\ Santiago, Chile 
       }

\date{}

\begin{document}

\maketitle

\begin{abstract}
In this article we present Zahir, a framework for experimentation in Computer Graphics that provides a group of object-oriented base components that take care of common tasks in rendering techniques and algorithms, specially those of Non Photo-realistic Rendering (NPR). These components allow developers to implement rendering techniques and algorithms over static and animated meshes. Currently, Zahir is being used in a Master's Thesis and as support material in the undergraduate Computer Graphics course in University of Chile.
\end{abstract}

\section{Introduction}

In this article we present a framework for Computer Graphics named \textit{Zahir}. This framework is currently in development and its lifespan covers a year and a half. It was conceived when we realized that we needed a tool to experiment with algorithms in Computer Graphics. 

The implementation of graphic techniques sometimes requires operations that are not part of the algorithm itself. This is natural, as the techniques can be very complex. When a developer needs to implement a technique, he has the following alternatives:
 \textit{1)} develop it from scratch, \textit{2)} use a graphics engine and provide only the minimal configuration and source code needed, and \textit{3)} use one or more libraries to support the development. The best choice depends on the context, for example, a student wants to learn how the technique works, so it would choose between the first and third option, while a developer for a commercial project needs a tested and fast solution like the second option.

Considering that situation, we propose Zahir as a framework for experimentation in Computer Graphics that provides a group of base components that take care of the most common tasks in techniques and algorithms, specially those of Non Photo-realistic Rendering (NPR). Having this group of base components allows developers to implement and experiment with the required techniques. 

\paragraph*{Contributions:}

The main contribution of our work is the design and implementation of Zahir, because it allows development of rendering techniques that are not implemented in common rendering engines, specially non photo-realistic rendering algorithms. Also, as Zahir is open source, it can be used as a tool by anyone.

\paragraph*{Outline.} This article is structured as follows: Section \ref{sec:related_work} discusses similar engines/frameworks. Section \ref{sec:zahir} contains the fundamental design and  implementations details. Section \ref{sec:usage} reviews the current usage of Zahir. Section \ref{sec:discussion} discusses who should use Zahir, limitations and future work, and finally Section \ref{sec:conclusions} shows our final conclusions for this article.

\section{Related Work}
\label{sec:related_work}

There are two types of frameworks/applications similar to Zahir: graphic engines and NPR applications.

\paragraph*{Graphic Engines.}
There are many rendering engines that can do traditional (photo-realistic) rendering in a optimal way (in terms of performance and image quality), such as \textit{OGRE}\footnote{\url{http://www.ogre3d.org}}, \textit{OpenSG}\footnote{\url{http://www.opensg.org/}} and \textit{OpenSceneGraph}\footnote{\url{http://www.openscenegraph.org/}}, all of them open source. They are mature solutions for rendering, with many years of development. They are very good solutions for a developer which needs to feed his data into the engine and just get them rendered. This is the common escenario, but in some cases that is not enough, as we may need adjacency and connectivity information (which may be available, but it's not as flexible or fast as we would like) that is not present in those engines. Also, they don't have mesh processing capabilities, or have very limited algorithms, thus limiting the posibilities in rendering techniques, specially those from NPR.

\paragraph*{Non Photo-realistic Rendering Applications.}
The inspiration for the initial version of Zahir was the software \textit{Real Time Suggestive Contours} \cite{decarlo2004scrt} (RTSC). RTSC renders line drawings from 3D surfaces in a beautiful and optimal way. However, because it's a demonstration program, it's limited to static white objects (the models don't load any material or texture information). Those limitations inspired us to start working on a more flexible framework.

Another similar tool is \textit{FreeStyle} \cite{GTDS04}. FreeStyle is a stylized line-renderer for 3D models. It's different from RTSC, because while RTSC extracts contours as iso-curves from surfaces, FreeStyle extracts silhouette and crease edges. The extracted edges are connected and stylized using procedures defined by the user. This software produces pretty pictures and is very customizable, but is not suited for interactive applications. On the other hand, it's very flexible and it can be used by artists because the styling can be written in the Python programming language.

\section{Zahir: Fundamentals}
\label{sec:zahir}

Zahir has the following goals:

\begin{itemize}
 \item \textit{Provide a high-level abstraction of the rendering pipeline}, by having a group of base components that take care of low-level tasks and implementation details.
 \item \textit{Ease the implementation of rendering techniques}, by implementing behavior common to many rendering techniques. 
 \item \textit{Complain with the following principles: efficiency, robustness and extensibility}, by using efficient and robust tools, and having an extensible design through object-orientation expressed in design patterns \cite{gamma1995dpe} and generic programming.
\end{itemize}

To reach these goals, we use libraries that share the same principles. The main library used by Zahir is \textit{OpenMesh} \cite{botsch2002oga}, which provides a \textit{half-edge} data-structure for polygonal meshes. We also integrate the matrix library \textit{Eigen2} \footnote{\url{http://eigen.tuxfamily.org/}}  and the numerical library \textit{OpenNL} \footnote{\url{http://alice.loria.fr/index.php/software.html}}. We use \textit{OpenGL} \cite{shreiner2003opg} as graphical API and \textit{Cg} \cite{fernando2003ctd} as shading language. This allows us to be portable across platforms and be compliant with the principles mentioned.

\subsection{Design Components}

\begin{figure}[htbp]
 \centering
 \includegraphics[width=0.6\textwidth]{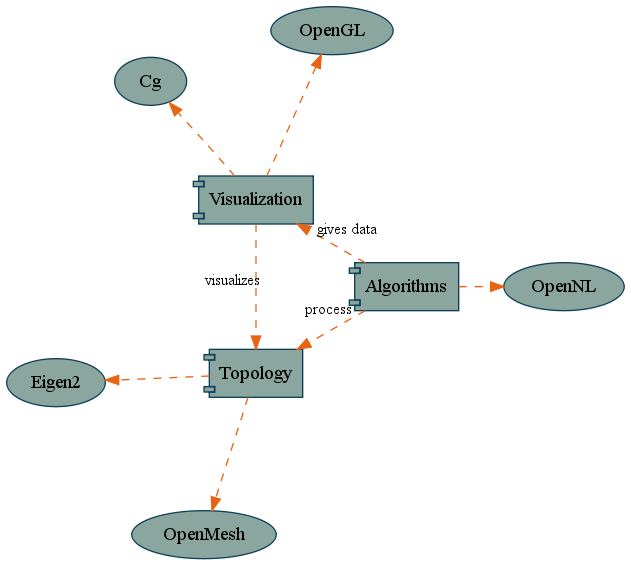}
 \caption{The base components of Zahir and the libraries used by them.}
 \label{fig:componentes}
\end{figure}

Zahir is composed by three components, shown on Figure \ref{fig:componentes}. The first one is the \textit{Topology Component}, which contains classes for static models, vertex animated models and skinned models through vertex blending. The second is the \textit{Algorithms Component}, which contains classes that process the mesh.  The third component is the \textit{Visualization Component}, which takes input from the previous components and displays them on screen. Each component uses different libraries: in Topology we need OpenMesh and Eigen2, both of them integrated (thanks to generic programming) to provide efficient and robust mesh data structures; in Algorithms we need OpenNL, because some operations require a numerical solver; in Visualization we need OpenGL and Cg to render the input (such as surfaces, lines and images) in 2D and 3D.

\subsection{Topology}

\begin{figure*}[htb]
 \centering
 \includegraphics[width=0.7\textwidth]{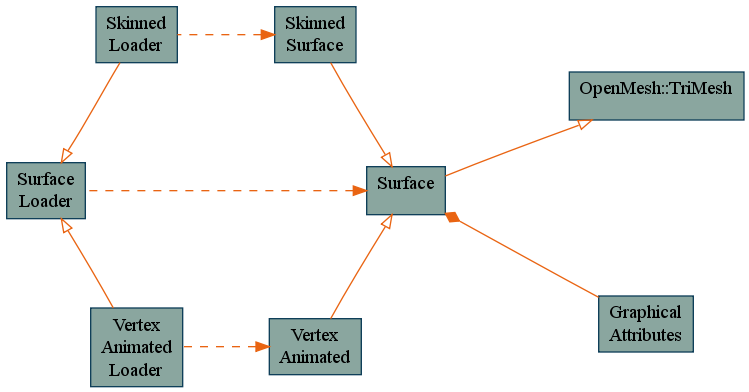}
 \caption{Classes of the Topology Component in Zahir.}
 \label{fig:componente_topologica}
\end{figure*}

Figure \ref{fig:componente_topologica} illustrates the design of the Topology Component. The main class, \texttt{Surface}, inherits from a triangular mesh in OpenMesh and has several graphic and mesh attributes such as material information, bounding sphere and feature size. The animated surfaces are a specialization of the \texttt{Surface} class, supporting two types of animation:

\begin{itemize}
 \item \textit{Vertex Animated Surfaces}, that have \textit{keyframes} that represent the mesh in different times. To animate (or deform) the mesh in a certain time, we interpolate the mesh from two keyframes. Each keyframe holds its own vertex positions and vertex normals, but the mesh structure is the same for all keyframes. Keyframes can be grouped in animations, and the interpolation between keyframes can be linear or cubic (using a Catmull-Rom interpolation).
 \item \textit{Skinned Surfaces}, or vertex blended meshes. We have only one mesh, which represents the surface in the \textit{bind-pose position}. This mesh has a skeleton associated, and when this skeleton moves, the mesh is deformed accordingly. The skeleton is represented as a hierarchical tree of bones, where each vertex in the mesh is associated to one or more bones. 
\end{itemize}

Both approaches share the same mesh structure and graphical attributes. Each subclass \textit{decorates} the main class with the extra attributes (such as keyframes or the skeleton itself). This class hierarchy implies that we see animated meshes as an extension of static meshes, where in a time $t$ we can assume that the mesh is static. This implies that before rendering or processing the mesh for time $t$ we need to calculate the mesh deformation. We compute the deformation on the CPU, which is slower than GPU methods but allows us to be more flexible at the implementation of rendering techniques.

The surface hierarchy is mirrored by a hierarchy of data loaders that define \textit{factory methods} and \textit{abstract factories} to build concrete surfaces. For example, a \textit{Wavefront OBJ} data loader is a implementation of the \texttt{SurfaceLoader} abstract factory, while a \textit{Quake II MD2} data loader is a implementation of the \texttt{VertexAnimatedLoader} abstract factory, which is also a subclass of \texttt{SurfaceLoader}. This design allows us to decouple the surface data from the process of reading the surface data from a data source.

\subsection{Algorithms}

\begin{figure*}[htb]
 \centering
 \includegraphics[width=0.65\textwidth]{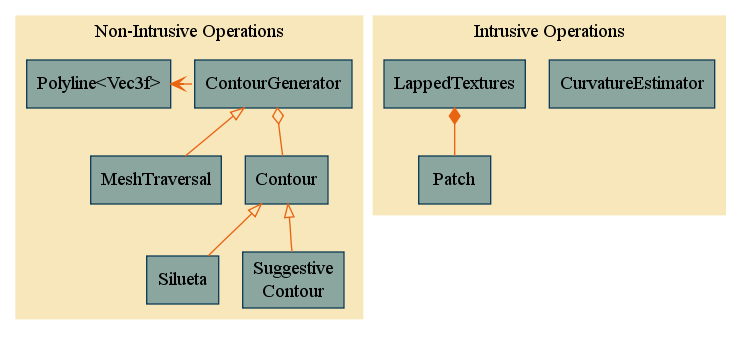}
 \caption{Classes of the Algorithms Component in Zahir.}
 \label{fig:componente_algoritmos}
\end{figure*}

The Algorithms Component, shown in Figure \ref{fig:componente_algoritmos} contains operations over the mesh. The operations that modify the mesh in some way are called \textit{intrusive}. Currently we have two intrusive operations: the estimation of curvature and curvature derivatives at each vertex of the mesh, based on the algorithm by Rusinkiewicz \cite{rusinkiewicz2004}, and \textit{Lapped Textures} \cite{praun2000lapped}, an algorithm that segments the mesh in patches and performs a local parameterization on each patch according to a vector field defined on the surface. Both of them work only on static meshes: their application on animated surfaces is a open problem.

The operations that extract information from the mesh without modifying it are called \textit{non-intrusive}. Currently, Zahir has one operation of this kind: the extraction of contours from a mesh, based on a contour definition (class \texttt{Contour}). We define a contour by specifying a function for evaluation at each vertex of the mesh. After the function is evaluated, we loop over the edges of the mesh finding the iso-curves of level $0$ by looking for edges whose vertices have opposite signs on the evaluation. If a edge has a level $0$ point, we iterate through its \textit{1-ring} of edges to seek more points of the same iso-curve. The extraction is specified by the interface \texttt{ContourGenerator} and implemented by the class \texttt{MeshTraversal}. A extracted contour is expressed as a set of polylines (class \texttt{Polyline}). 

We have two contours defined in Zahir: \textit{silhouettes}\cite{lake2000stylized} and \textit{Suggestive Contours} \cite{decarlo2003}. Both are view-dependant contours, which means that after changes in the viewpoint or in the 3D scene the contours must be extracted again. 

\subsection{Visualization}

\begin{figure*}[htb]
 \centering
 \includegraphics[width=0.6\textwidth]{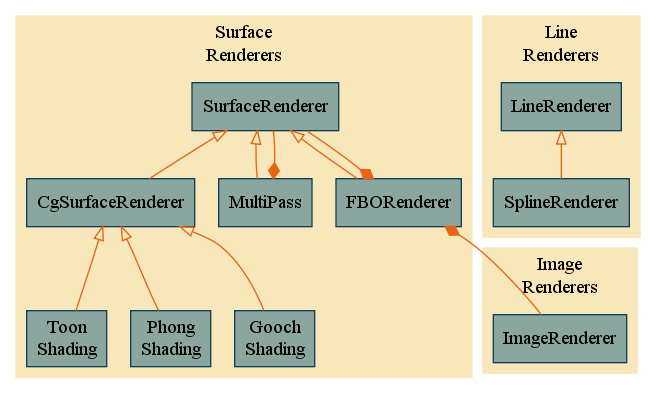}
 \caption{Classes of the Visualization Component in Zahir.}
 \label{fig:componente_graficacion}
\end{figure*}

The Visualization Component, shown in Figure \ref{fig:componente_graficacion}, takes input from the previous components to produce an image on the screen using OpenGL and Cg. The classes contained in this component can be grouped in three kinds of renderers: \textit{surface renderers}, \textit{line renderers} and \textit{image renderers}. 

\subsubsection{Surface and Image Renderers}

The \texttt{SurfaceRenderer} is the base class for the group of surface renderers. It defines a basic OpenGL renderer for the base class of the Topology Component, \texttt{Surface}. Because at the time of rendering we treat any surface as a static one, for animated surfaces we need an update to the mesh (calculate the deformation) previously to rendering.

The class hierarchy follows two design patterns. The first one is \textit{decorator}, because each subclass decorates its parent class with rendering parameters and options. Since the rendering pipeline is almost independent from the rendering technique or style, this behavior is attractive for the developer, as the only thing he needs to do to create new techniques is to specify the parameters and properties of the technique. This is specially flexible thanks to the surface properties mechanism of OpenMesh.

The other pattern is \textit{composite}. We needed a way to treat single-pass and multi-pass techniques in the same way. Since sometimes a pass in a multi-pass technique is already a single-pass technique,  this design allows us to create small renderers that can be assembled to create more complex ones.

\begin{figure*}[htb]
\begin{center}
\subfigure[Toon Shading]{ \includegraphics[width=0.31\textwidth]{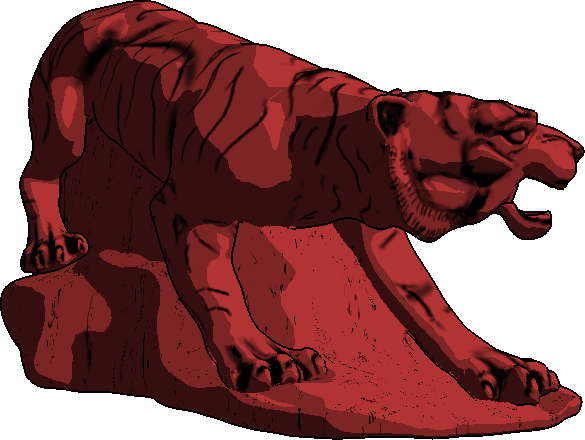}}
\subfigure[Toon Shading]{ \includegraphics[width=0.31\textwidth]{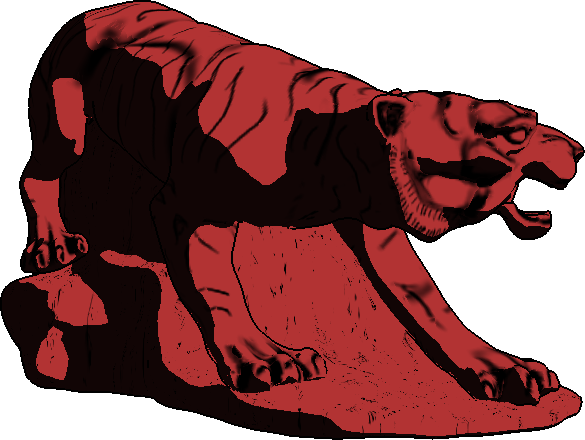}}
\subfigure[Gooch Shading]{ \includegraphics[width=0.31\textwidth]{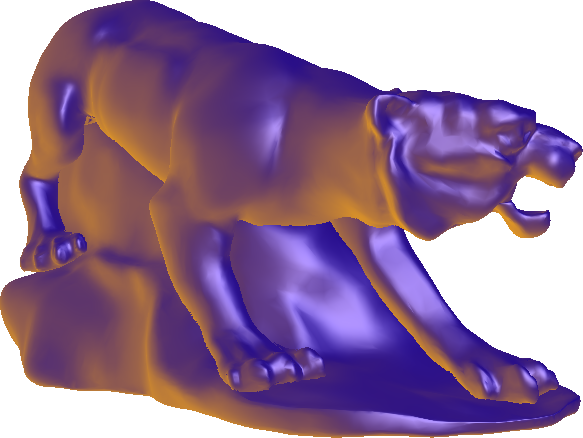}}
\caption{Surface renderers applied to a static mesh.}
\label{fig:graficadores_modelo}
\end{center}
\end{figure*}

Currently, we have the following techniques implemented: \textit{Phong Shading}, \textit{Gooch Shading} \cite{gooch1998} and \textit{Toon Shading} \cite{lake2000stylized}, shown in Figure \ref{fig:graficadores_modelo}. There is also a render to texture renderer (\texttt{FBORenderer}), that receives another surface renderer and stores its result in a texture.

The group of image renderers contains only one class, \texttt{ImageRenderer}. This class takes a input texture and draws a quad with the texture applied. Optionally, it can apply a \textit{fragment program} using Cg. This allows us to implement image processing techniques or image space algorithms when used in conjunction with the \texttt{FBORenderer}.

\subsubsection{Line Renderers}

\begin{figure*}[htb]
\begin{center}
\subfigure[No Interpolation] { \includegraphics[width=0.31\textwidth]{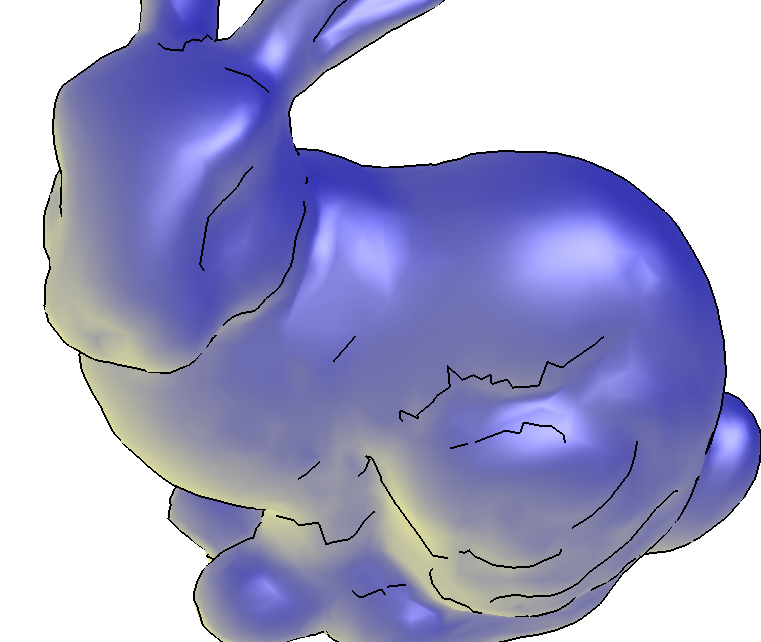}}
\subfigure[Catmull-Rom]{ \includegraphics[width=0.31\textwidth]{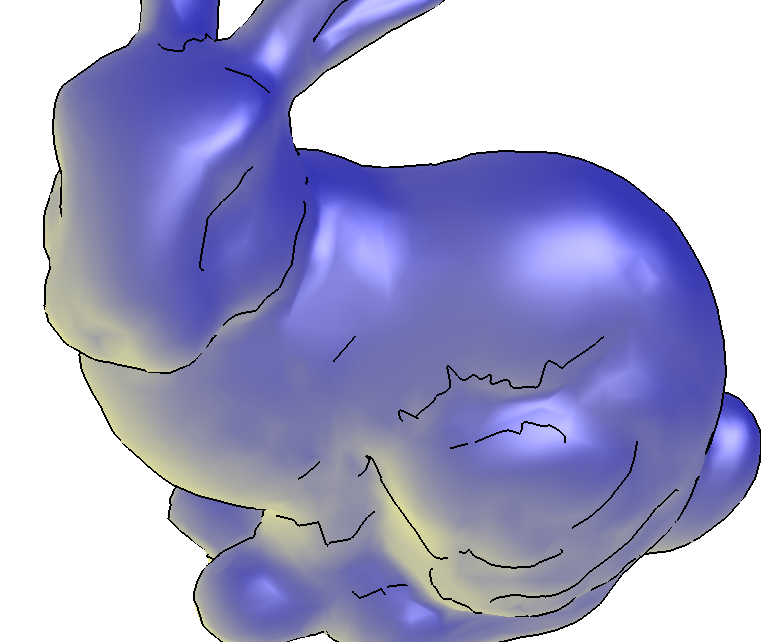}}
\subfigure[B-Spline]{ \includegraphics[width=0.31\textwidth]{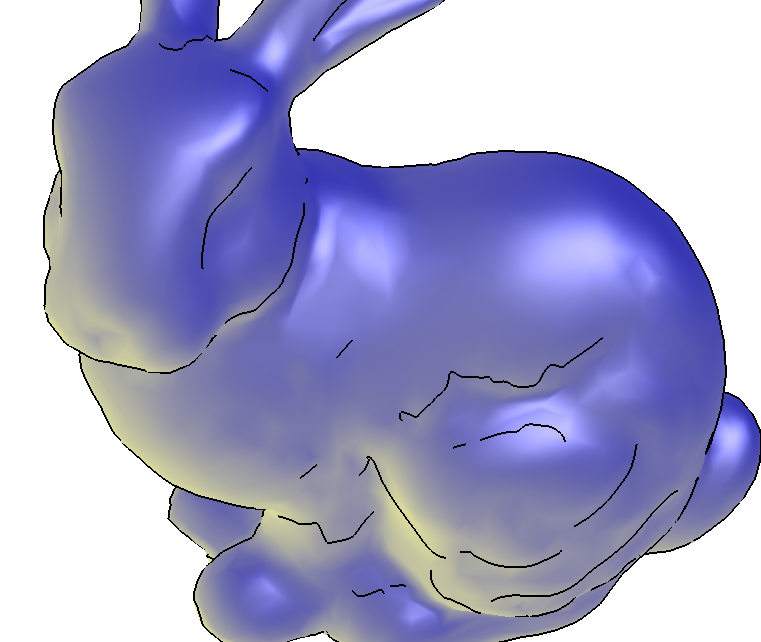}}
\caption{Gooch Shading with Silhouettes and Suggestive Contours. The three images show Suggestive Contours with different line renderers: the base renderer (left), the spline renderer with Catmull-Rom interpolation (center), the spline renderer with B-Spline interpolation (right).}
\label{fig:graficadores_lineas}
\end{center}
\end{figure*}

The group of line renderers take the extracted contours by the Algorithms Component and displays them on screen. It has two classes: \texttt{LineRenderer}, the base class, and \texttt{Spline}, a subclass that decorates the base renderer with a cubic interpolator. Figure \ref{fig:graficadores_lineas} shows a coarse mesh with silhouettes and suggestive contours extracted. The Figure shows how different suggestive contours look by using direct polyline rendering (left), Catmull-Rom interpolation (center) and B-Spline interpolation (right). 

\subsection{Rendering Loop}

\begin{figure*}[htb]
 \centering
 \includegraphics[width=0.9\textwidth]{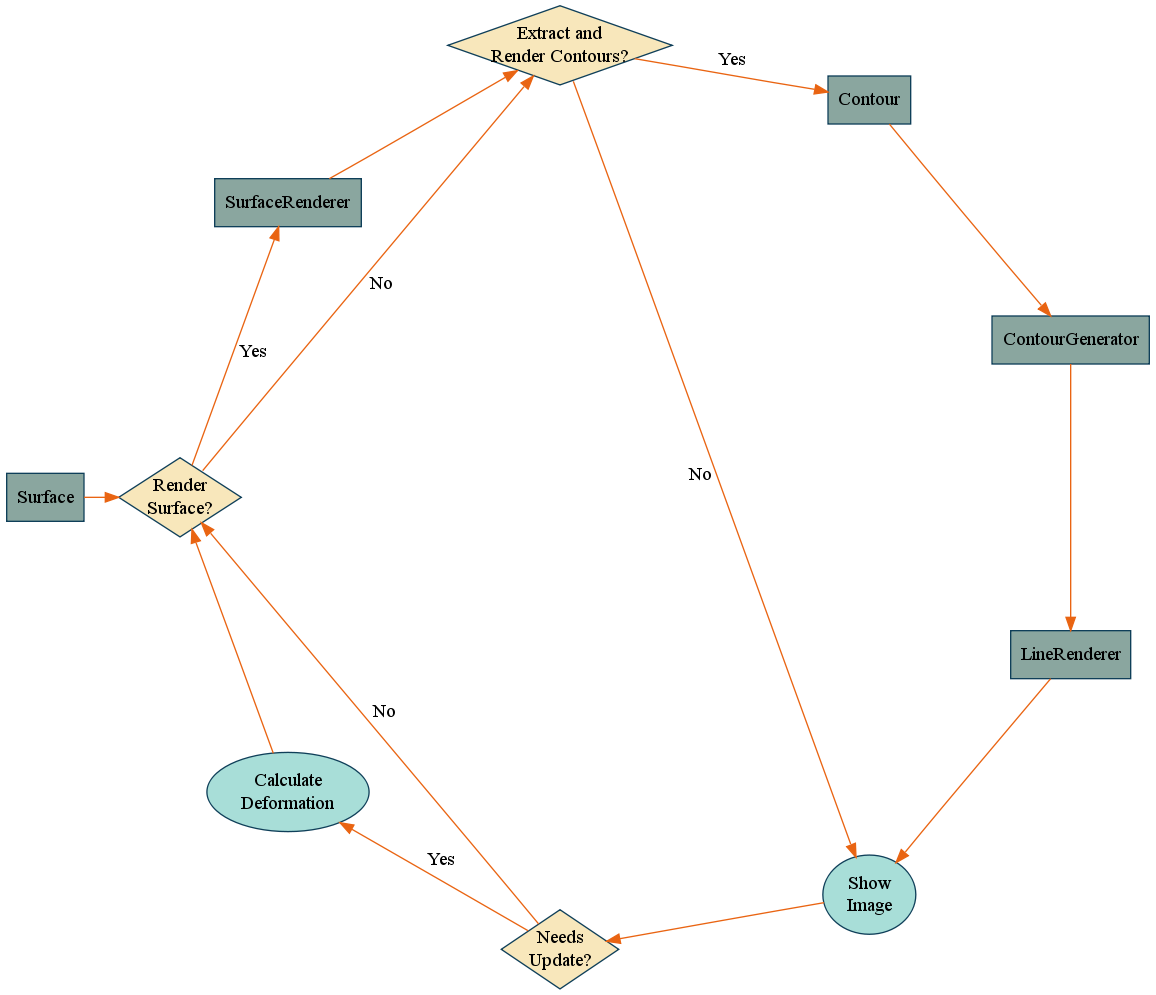}
 \caption{Rendering loop for an arbitrary Surface, considering both surface and line renderers.}
 \label{fig:flujo_graficacion}
\end{figure*}

Figure \ref{fig:flujo_graficacion} shows a rendering loop for an arbitrary surface in a typical application. We first take a input mesh, select a surface renderer if needed, then select the contours to extract and render both surface and lines. Finally, before starting the loop again, update the mesh (according to user input and time passed) if needed. As can be seen in the diagram, each step refers to the base class of its corresponding group. In this way, we can add new behaviour just by adding subclasses, allowing more flexibility in the application at the cost of runtime polymorphism.

\section{Current Usage}
\label{sec:usage}

A framework needs users to fill its purpose. A framework without users doesn't make sense, and that's why we are encouraging the use of Zahir in our university. The current usage for our framework can be divided in experimentation and teaching.

\subsection{Experimentation}

One of the purposes of Zahir is the experimentation with rendering techniques and mesh algorithms. This experimentation includes the implementation of known techniques and the investigation of new ones. Currently we are implementing a non photo-realistic rendering technique called \textit{Realtime Hatching} \cite{praun2001hatching} in a Master's Thesis. Although there is no real experimentation by just implementing the hatching, by finishing its implementation we will have the possibility to experiment with the algorithm in animated surfaces, something that hasn't been done in object-space yet.

\subsection{Teaching}

\begin{figure*}[htb]
 \centering
 \subfigure{ \includegraphics[height=0.15\textheight]{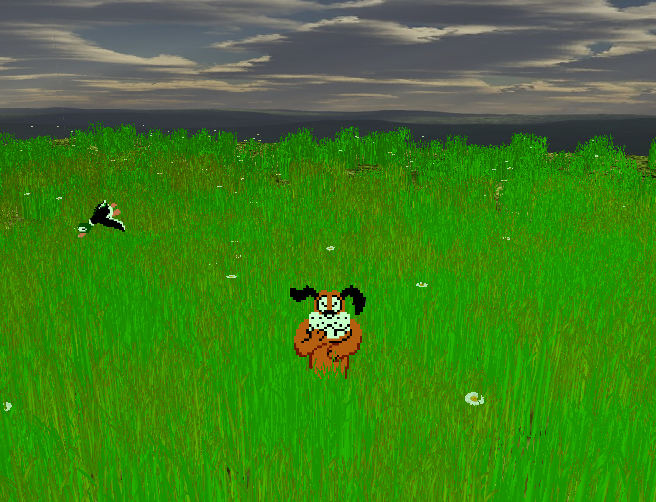}}
 \subfigure{ \includegraphics[height=0.15\textheight]{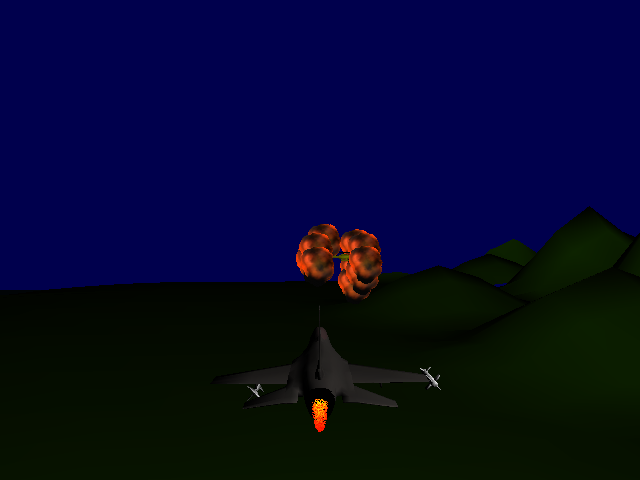}}
 \subfigure{ \includegraphics[height=0.15\textheight]{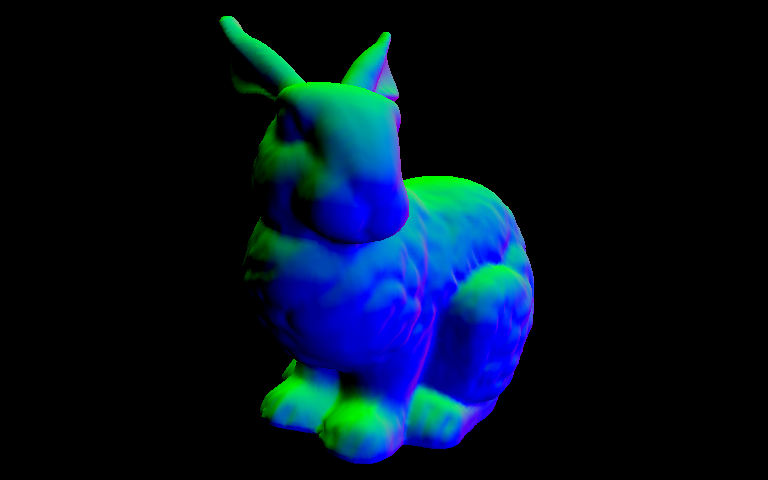}}
 \label{fig:proyectos}
 \caption{Example projects from undergraduate students of the Computer Graphics course in University of Chile.}
\end{figure*}

During the year 2008 we have used Zahir as a optional support tool for student projects and homeworks in the Computer Graphics undergraduate course in University of Chile. The response has been great, as some students used the framework as a basis for their projects. Since Zahir implements many concepts of the course program, it serves as a reference implementation for many concepts even if the students decide not to use it as a main library in their projects. Figure \ref{fig:proyectos} shows three example projects from the Spring semester in 2008. The first (left) used the interpolation classes of Zahir, the second (center) used the surface and image renderers to display a simple terrain and a combat aeroplane, and the third (right) implemented different renderers such as normal map and depth map to apply border detection in image-space. 

\section{Discussion}
\label{sec:discussion}

In Section \ref{sec:zahir} we mentioned similar applications. Should we use Zahir instead of them? The answer depends on the user, but probably they are better alternatives in a typical scenario (to develop a computer graphics application) considering how young Zahir is. However, there's a niche of opportunity in some use cases, mainly in those related to non photo-realistic rendering and experimentation. This is because common rendering engines don't provide means to perform operations on meshes, establishing limits on the implementation of object-space techniques because of the lack of other information than vertex positions, normals and materials.

\paragraph*{Who should use Zahir?}

Zahir should be used by computer graphics students or enthusiasts seeking for a base library to work. We have designed Zahir in a way to be flexible for many different uses, in particular experimentation, because it allows the users to work directly in their ideas and problems without obstacles like object loading from file and rendering setup.

\paragraph*{Limitations and Future Work:}

Apart from the typical limitations of new software libraries (like lack of documentation and code examples), Zahir has a important limitation that can be a barrier for some users: it is useful to implement rendering techniques and mesh algorithms, but not to specify them using a script language or configuration files. Also, while we support three kinds of surfaces, currently we don't support many filetypes or formats to load data from external programs. 

Future work should address these limitations, by creating renderers that take input from configuration files (possibly in XML) and data loaders for formats such as COLLADA or Blender. In terms of functionality and features, we plan to put emphasis on the efficiency principle of Zahir: the libraries included in the framework are efficient by themselves, but to operations using them we have given priority to robustness and flexibility. Also, in the future we plan to move some calculations to the GPU to improve performance in operations such as contour extraction.

\section{Concluding Remarks}
\label{sec:conclusions}

We have presented a framework for computer graphics named Zahir. Zahir is a young framework with many limitations to adress in the future, but we have already identified use cases where it stands well. Zahir has an object oriented design, expressed through known design patterns, which makes it flexible to add new behavior such as rendering techniques and operations over meshes. Since we were inspired by non photo-realistic rendering tools, our work is biased to this kind of applications, but we also support traditional rendering.

We have tested the framework with undegraduate students, having good results. This approach has allowed us to receive feedback from potentially ``early adopters'' and, when their projects end well, add their contributions to the framework, making it more feature complete and allowing it to grow.

\bibliographystyle{plain}
\bibliography{zahir}

\end{document}